\documentclass[12pt]{article}
 \usepackage{amsmath,amssymb,amscd,amsthm}
\usepackage{latexsym}
\usepackage[dvips]{graphics,graphicx,epsfig}
\usepackage{graphicx}

 \numberwithin{equation}{section}
 \newtheorem{thm}{Theorem}[section]
 \newtheorem{lem}{Lemma}[section]
 \newtheorem{prop}{Proposition}[section]
 \theoremstyle{definition}
 
 \def\R{\mathbb{R}}
 
 \def\Z{\mathbb{Z}}

 \title{ Stability of periodic
traveling waves for the quadratic and cubic nonlinear Schr\"odinger  equations}
\author{Sevdzhan Hakkaev$ ^1$, Iliya D. Iliev$ ^2$  and
   Kiril Kirchev$ ^2$ }

\begin{document}

\maketitle

\begin{center}
{$ ^1$Faculty of Mathematics and Informatics,

 Shumen University, 9712 Shumen, Bulgaria

 \vspace{2ex}
 $ ^2$Institute of Mathematics and Informatics,

 Bulgarian Academy of Sciences, 1113 Sofia, Bulgaria \\ }
 \end{center}

\begin{abstract} We study the existence and stability of
periodic traveling-wave solutions for the quadratic and cubic
nonlinear Schr\"odinger equations in one space dimension.
\end{abstract}

\section{Introduction}
   In this work we consider the nonlinear Schr\"odinger equation
   (NLS)
     \begin{equation}\label{1.1}
       iu_t+u_{xx}+|u|^pu=0.
     \end{equation}
     This equation appears in various problems, modeling many
     phenomena such as the behavior of a non-ideal Bose gas with a
     weak particle interaction, the spreading of the heat impulse
     in solids, the Langmuir waves in a plasma, etc. \cite{Wh,
     Za}.

   Our principal aim is to study the the orbital stability of the family of periodic
  traveling-wave solutions
\begin{equation}\label{1.2}
u=\varphi(x,t)=e^{i(vx+(\omega-v^2)t)}r(x-2vt).
\end{equation}
where $r(y)$ is a real-valued $T$-periodic function and $v,
\omega\in\R$ are parameters,
   for quadratic (p=1) and cubic (p=2)
   nonlinear Schr\"odinger equation.The problem of the stability of
solitary waves for nonlinear dispersive equations goes back to the
works of Benjamin \cite{Be1} and Bona \cite{Bo} (see also
\cite{AlBoHe, W1, W2}). A general approach for investigating the
stability of solitary waves for nonlinear equations having a group
of symmetries was proposed in \cite{GSS}. The existence and
stability of solitary wave solutions for equation (\ref{1.1}) has
been studied in \cite{ZIK}.  Recently in \cite{ABS}, the authors
developed a complete theory on the stability of cnoidal waves for
the KdV equation. Other new explicit formulae for the periodic
traveling waves based on the Jacobi elliptic functions, together
with their stability, have been obtained in \cite{An2, HIK1, HIK2}
for the nonlinear Schr\"odinger equation, modified KdV equation,
complex modified KdV equation, and generalized BBM equation. In
\cite{Ha}, the stability of periodic traveling wave solutions of
BBM equation which wave profile stays close to the constant state
$u=(c-1)^{1/p}$ is considered.

In this paper, we prove stability of the periodic traveling waves
(\ref{1.2}) not oscillating around zero $(r\neq 0)$
for the quadratic and the cubic NLS. Our main results are formulated
as Theorem \ref{t21} and Theorem \ref{t31} in Sections 2 and 3 below.
  We base our analysis on some appropriate invariant laws.
  Our approach is to verify that $\varphi$ is a
minimizer of a properly chosen functional $M$ which is
conservative with respect to time over the solutions of
(\ref{1.1}). We consider the $L^2$-space of $T$-periodic functions
in $x\in\R$, with a norm $||.||$ and a scalar product $\langle
.,.\rangle$. To establish that the orbit
$${\cal O}=\{e^{i\eta}\varphi(\cdot-\xi,t):\;
(\xi,\eta)\in[0,T]\times[0,2\pi]\}$$
is stable, we take
$$u(x,t)=e^{i\eta}\varphi(x-\xi,t)+h(x,t)=e^{i\zeta}[r(x-\xi-2vt)+h_1+ih_2]$$
and express the leading term of $M(u)-M(\varphi)$ as $\langle
L_1h_1,h_1\rangle+\langle L_2h_2,h_2\rangle$ where $L_i$ are
second-order selfadjoint differential operators in $L^2[0,T]$ with
potentials depending on $r$ and satisfying $L_1 r'=L_2r=0$. The
proof of orbital stability requires that zero is the second
eigenvalue of $L_1$ and the first one of $L_2$.

Recall that the  quadratic and cubic nonlinear Schr\"odinger equations are
globally well-posed in $H^s(\mathbb{T})$, for $s\geq 0$
\cite{Bour}.

The paper is organized as follows. We consider the quadratic and the cubic
cases in Sections 2 and 3 respectively. In Appendices 1 and 2, some facts
from the theory of complete Abelian integrals (e.g. Picard-Fuchs equations,
polynomial moduli) are used in order to derive several inequalities we
needed during the proof our main results.

%%%%%%%%%%%%%%%%%%%%%%%%%%%%%%%%%%%%%%%%%%%%%%%%%%%%%%%%%%%%%%%%%%%%%%%%

\section{Existence and stability of periodic traveling waves for the
quadratic Schr\"odinger equation}

Consider the equation
    \begin{equation}\label{2.1}
    iu_{t}+u_{xx}+|u|u=0,
   \end{equation}
  where $u$ is a complex-valued function.

  We are looking for a solution of equation (\ref{2.1}) in the form
  (\ref{1.2}) where $r$ is real-valued. For $r$ one obtains the equation
    \begin{equation}\label{2.2}
    r''-\omega r +r|r|=0.
   \end{equation}
Therefore,
    \begin{equation}\label{2.3}
    r'^2-\omega r^2 +\frac23r^2|r|=c
   \end{equation}
and $r$ is periodic provided that the level set $H(x,y)=c$ of the
Hamiltonian system $dH=0$, with
$$H(x,y)=y^2-\omega
x^2+\frac23x^2|x|,$$ contains a periodic trajectory (an oval). The
level set $H(x,y)=c$ contains two periodic trajectories if
$\omega>0$, $c\in(-\frac13\omega^3,0)$ and a unique periodic
trajectory if $\omega\in\R$, $c>0$. Under these conditions,
equation (\ref{2.3}) becomes $H(r,r')=c$ and its solution $r$ is
periodic of period $T=T(\omega,c)$.

Below, we consider the case $c<0$. Then either $r<0$ (the left
case) or $r>0$ (the right case). To express $r$ through elliptic
functions, we denote by $r_0>r_1>0$ the positive solutions of
$\frac23\rho^3-\omega \rho^2-c=0$. Then $r_1\leq |r|\leq r_0$ and
one can rewrite (\ref{2.3}) as
\begin{equation}\label{2.4}
\textstyle
r'^2=\frac23(|r|-r_1)(r_0-|r|)(|r|+r_0+r_1-\frac32\omega).
\end{equation}
Therefore $2r_0+r_1>r_0+2r_1>\frac32\omega$. Introducing a new
variable $s\in(0,1)$ via $|r|=r_1+(r_0-r_1)s^2$, we transform
(\ref{2.4}) into $$s'^2=\alpha^2(1-s^2)(k'^2+k^2s^2)$$ where
$\alpha$, $k$, $k'$ are positive constants ($k^2+k'^2=1$) given by
$$\alpha^2=\frac{4r_0+2r_1-3\omega}{12}, \quad
k^2=\frac{2r_0-2r_1}{4r_0+2r_1-3\omega}, \quad
k'^2=\frac{2r_0+4r_1-3\omega}{4r_0+2r_1-3\omega}.$$ Therefore
\begin{equation}\label{2.5}
|r(x)|=r_1+(r_0-r_1)cn^2(\alpha x;k).
\end{equation}

\begin{equation}\label{2.51}
  T={\frac{2K(k)}{\alpha}}={\frac{4\sqrt[4]{1-k^2+k^4}K(k)}{\sqrt{\omega}}}, \;  k     \in (0,1), \; T\in I=\left( {\frac{2\pi}{\sqrt{\omega}}}, \infty \right).
\end{equation}
As usual, here and below, $K(k)$ and $E(k)$ denote the complete elliptic
integrals of the first and the second kind in a Legendre form.
Let us recall for later use the system they satisfy:
$$kK'=\frac{E}{1-k^2}-K,\;\; kE'=E-K.$$

\vspace{1ex} \noindent \begin{lem}\label{l22}  For any $\omega>0$
and $T\in I$, there is a constant $c=c(\omega)$ such that the
periodic traveling-wave solution $(\ref{2.5})$ determined by
$H(r,r')=c(\omega)$ has a period $T$. The function $c(\omega)$ is
differentiable.
\end{lem}

\vspace{2ex}
\noindent
{\bf Proof.} The statement follows from the implicit function theorem.
It is easily seen that the period $T$ is a strictly increasing function of $k$:
$$\begin{array}{ll}
\frac{d}{dk}(\sqrt[4]{1-k^2+k^4}K(k))&=
\frac{k(2k^2-1)K(k)+2(1-k^2+k^4)K'(k)}
{2(1-k^2+k^4)^{3/4}}\\
\\
&=\frac{2(1-k^2+k^4)E(k)+(1-k^2)(k^2-2)K(k)}{2k(1-k^2)(1-k^2+k^4)^{3/4}}>0.
\end{array}$$
Given $\omega$ and $c$ in their range, consider the functions $r_0(\omega,c), r_1(\omega,c)$,
$k(\omega,c)$ and $T(\omega,c)$ given by the formulas we derived above.
We obtain
$$\frac{\partial T}{\partial c}=\frac{dT}{dk} \frac{dk}{dc}=
\frac{1}{2k}\frac{dT}{dk}\frac{d(k^2)}{dc}.$$
Further, using that $k^2=3{\frac{2r_0-\omega}{4r_0+2r_1-3\omega}}-1$ and
${\frac{2}{3}}r_0^3-\omega r_0^2={\frac{2}{3}}r_1^3-\omega r_1^2$, we have
$$\begin{array}{ll}\displaystyle
\frac{d(k^2)}{dc}&\displaystyle=3{\frac{(4r_1-2\omega){\frac{\partial r_0}
{\partial c}}-(4r_0-2\omega){\frac{\partial r_1}{\partial c}}}{(4r_0+2r_1
-3\omega)^2}}\\[2mm]
&\displaystyle={\frac{3\omega^2(r_1-r_0)}
{4(r_0^2-\omega^2)(r_1^2-\omega^2)(4r_0+2r_1-3\omega)^2}}.
%
%{\frac{\omega(r_0-r_1)}{4(r_0^2-\omega r_0)(r_1^2-\omega r_1)
%(4r_0+2r_1-3\omega)^2}}(5r_0+5r_1-4\omega).
%
\end{array} $$
We see that $\partial T(\omega,c)/\partial c\neq 0$, therefore
the implicit function theorem yields the result. $\Box$

\vspace{2ex}
Equation (\ref{2.1}) has the following conservation laws
  $$
     Q(u)=i\int_{0}^{T}{\overline{u}_{x}u}dx,\quad
     P(u)=\int_{0}^{T}{|u|^{2}}dx,\quad
     E(u)=\int_{0}^{T}{(|u_{x}|^{2}-{\frac{2|u|^{3}}{3}})}dx.
  $$
  Let us consider the functional
    $$M(u)=E(u)+(\omega +v^{2})P(u)-2vQ(u).$$
    Next we introduce the pseudometric
  \begin{equation}\label{2.5a}
  d(u, \varphi)=\inf_{(\eta, \xi)\in [0, 2\pi]\times [0,T]}
  ||u(x,t)-e^{i\eta}\varphi(x-\xi, t)||_{1}.
  \end{equation}
    For a fixed $q>0$, we denote
     \begin{equation}\label{2.5b}
      d_{q}^{2}(u, \varphi)= \inf_{(\eta, \xi)\in [0, 2\pi]\times
     [0,T]} (||u_{x}(x,t)-e^{i\eta}\varphi_{x}(x-\xi, t)||^2
     +q||u(x,t)-e^{i\eta}\varphi(x-\xi, t)||^2).
     \end{equation}
Clearly, the infimum in  $(\ref{2.5b})$ is attained at some point
$(\eta, \xi)$ in the square $[0,T]\times [0,T]$. Moreover, for
$q\in [q_1,q_2]\subset(0,\infty)$, (\ref{2.5b}) is a pseudometric
equivalent to (\ref{2.5a}).

\begin{lem}\label{l21}
The metric $d_{q}(u,\varphi)$ is a continuous function of $t\in
[0, \infty)$.
      \end{lem}

\noindent {\bf Proof.} The proof of the lemma is similar to the proof of
Lemmas 1, 2 in \cite{Bo} $\Box$.

  Now, we can formulate our main result in this section.

\begin{thm}\label{t21}
Let $\varphi$ be given by $(\ref{1.2})$, with $r\neq 0$.
         For each $\varepsilon>0$ there exists $\delta>0$ such that if
         $u(x,t)$ is a solution of $(\ref{2.1})$ and
         $d(u, \varphi )_{|t=0}<\delta$, then $d(u, \varphi)<\varepsilon$
         $\forall t\in [0,\infty)$.
       \end{thm}

The crucial step in the proof will be to verify the following
statement.

\begin{prop}\label{p21}
  There exist positive constants $m, q, \delta_{0}$ such that if $u$ is
a periodic solution of $(\ref{2.1})$,  $u(x,t)=u(x+T,t)$,
$P(u)=P(\varphi)$ and $d_{q}(u, \varphi)<\delta_{0}$, then
    \begin{equation}\label{2.6}
   M(u)-M(\varphi)\geq m d_{q}^{2}(u, \varphi).
    \end{equation}
 \end{prop}

 \vspace{1ex}
\noindent
{\bf Proof.} In order to estimate $\Delta M=M(u)-M(\varphi)$, we
set
   $$u(x,t)=e^{i\eta}\varphi(x-\xi, t)+h(x,t),$$
   $$\zeta =v(x-\xi)+(\omega-v^2)t+\eta ,\;\;
   F(s)=-{\frac{2}{3}}|e^{i\eta}\varphi+hs|^3.$$
   We have
$$
\frac23(-|u|^3+|\varphi|^3)=F(1)-F(0)
=F'(0)+\frac{F''(0)}{2}+\frac{F''(s)-F''(0)}{2}
$$
     where
$$
  0\leq s \leq 1, \; F'(0)=-2|\varphi| Re(e^{i\eta} \varphi \overline{h}),\;
  F''(0)=-|\varphi|\left[Re(h^2e^{-2i\zeta})+3|h|^2\right].
$$
   Integrating by parts in the terms containing $h_{x}$ and
   $\overline{h}_{x}$, we obtain
     $$\begin{array}{ll}
        \Delta M&=M(u)-M(\varphi) \\
      \\
      &=2Re\int_0^T{e^{i\eta}[-\varphi_{xx}
      +(\omega+v^2-|\varphi|)\varphi+2iv\varphi_x]\overline{h}}dx\\
      \\
      &+\int_0^T{[|h_{x}|^2+(\omega+v^2-{\frac{3}{2}}|\varphi|)|h|^2
      -{\frac{|\varphi|}{2}}Re(e^{-2i\zeta}h^2)-2ivh\overline{h}_x]}dx\\
      \\
      &-\int_0^T{\frac{F''(s)-F''(0)}{2}}dx\\
      \\
      &=I_{1}+I_{2}+I_{3}.
        \end{array}
     $$
Using that $r(x)$ satisfies the equation (\ref{2.2}) we obtain
that
   $I_{1}=0$.

   Let $h(x,t)=(h_{1}+ih_{2})e^{i\zeta},$
   where $h_{1}$ and $h_{2}$ are real periodic functions with period $T$.
  Then we have
     \begin{equation}\label{2.7}
     \begin{array}{ll}
        |h|^{2}=h_{1}^{2}+h_{2}^{2}\\
        \\
        |h_{x}|^{2}=h_{1x}^{2}+h_{2x}^{2}+2v(h_{1}h_{2x}-h_{1x}h_{2})
        +v^{2}(h_{1}^{2}+h_{2}^{2})\\
        \\
        Re(e^{-2i\zeta}h^{2})=h_{1}^{2}-h_{2}^{2}\\
        \\
        \int_0^T {h\overline{h}_x}dx=i\int_{0}^{T}{(h_{1x}h_2
        -h_{1}h_{2x}-vh_{1}^{2}-vh_{2}^{2})}dx.
       \end{array}
     \end{equation}
     Finally, for $I_2$ we obtain
   $$
          I_{2}=\int_0^T{[h_{1x}^2+(\omega-2|r|)h_1^2 ]}dx
          +\int_0^T{[h_{2x}^2 +(\omega -|r|)h_2^2]}dx=M_1+M_2
       $$
Consider in $[0,T]=[0, 2K(k)/\alpha]$ the formal differential
operators $$L_1=-\frac{d^2}{dx^2}+(\omega-2|r|), \quad
L_2=-\frac{d^2}{dx^2}+(\omega-|r|),$$ supplied with periodic
boundary conditions. By the above formulas,
$r_0-r_1=6\alpha^2k^2$,
 $2r_0-\omega=4\alpha^2(1+k^2)$. Taking $y=\alpha x$ as an
independent variable in $L_1$, one obtains $L_1=\alpha^2\Lambda_1$
with an operator $\Lambda_1$  in $[0,2K(k)]$ given by
$$\begin{array}{rl} \Lambda_1&=\displaystyle -\frac{d^2}{dy^2}
+\alpha^{-2}[\omega-2(r_1+(r_0-r_1)cn^2(y;k))]\\[5mm]
&\displaystyle= -\frac{d^2}{dy^2}+
\frac{\omega-2r_0}{\alpha^2}+\frac{2(r_0-r_1)}{\alpha^2}sn^2(y;k)\\[5mm]
&\displaystyle= -\frac{d^2}{dy^2}-4(1+k^2)+12k^2sn^2(y;k).
\end{array}$$
The spectral properties of the operator  $\Lambda_1$  in
$[0,2K(k)]$ are well known. The first three(simple) eigenvalues
and corresponding eigenfunctions of $\Lambda_1$ are
  $$\begin{array}{ll}
    \mu_0=k^2-2-2\sqrt{1-k^2+4k^4}<0, \\
    \psi_0(y)=dn(y;k)[1-(1+2k^2-\sqrt{1-k^2+4k^4})sn^2(y;k)]>0\\
    \\
    \mu_1=0\\
    \psi_1(y)=dn(y;k)sn(y;k)cn(y;k)={\frac{1}{2}}{\frac{d}{dy}}cn^2(y;k)\\
    \\
    \mu_2=k^2-2+2\sqrt{1-k^2+4k^4}>0\\
    \psi_2(y)=dn(y;k)[1-(1+2k^2+\sqrt{1-k^2+4k^4})sn^2(y;k)].
   \end{array}
  $$
  Since the eigenvalues of $L_1$ and $\Lambda_1$ are related by
  $\lambda_n=\alpha^2 \mu_n$, it follows that the first three eigenvalues of the operator
      $L_1$, equipped with periodic boundary condition on $[0,2K(k)]$
      are simple and $\lambda_0<0, \lambda_1=0, \lambda_2>0$. The
      corresponding eigenfunctions are $\psi_0(\alpha x),
      \psi_1(\alpha x)=const. r'$ and $\psi_2(\alpha x)$.

In a similar way, with $L_2=\alpha^2\Lambda_2$, one obtains in
$[0, 2K(k)]$
$$
\Lambda_2=-\frac{d^2}{dy^2}-2(1+k^2)+6k^2sn^2(y;k)+\omega/2\alpha^2.$$
To express $\omega$ through $\alpha$ and $k$, one should take into
account  the fact that in the cubic equation we used to determine
$r_0$ and $r_1$,
 the coefficient at $\rho$ is zero. Therefore,
$$\textstyle r_0r_1+(r_0+r_1)(\frac32\omega-r_0-r_1)=0.$$ As
$r_0=2\alpha^2+2\alpha^2k^2+\frac12\omega$,
$r_1=2\alpha^2-4\alpha^2k^2+\frac12\omega$,  after replacing these
values in the above equation one obtains
$\omega^2=16\alpha^4(1-k^2+k^4)$. Since $\omega>0$, we finally
obtain
$$\Lambda_2=-\frac{d^2}{dy^2}+2(-1-k^2+\sqrt{1-k^2+k^4})+6k^2sn^2(y;k).$$
On the other hand, (\ref{2.5}) yields
$$|r|=2\alpha^2[1+k^2+\sqrt{1-k^2+k^4}-3k^2sn^2(y;k)].$$
 The first three eigenvalues and corresponding eigenfunctions
 of $\Lambda_2$ are as follows:
$$\begin{array}{ll} \epsilon_0=0, & \upsilon_0=r,\\[2mm]
 \epsilon_1=2-k^2+2\sqrt{1-k^2+k^4}, & \upsilon_1=dn'(y;k)\\[2mm]
  \epsilon_2=4\sqrt{1-k^2+k^4}, & \upsilon_2=1+k^2-\sqrt{1-k^2+k^4}-3k^2sn^2(y;k).
 \end{array}$$

\vspace{1cm}
\noindent
 {\bf Estimates for $M_2$.}

\vspace{0.5cm}
\noindent
 From the above explanations we know that when considered in
$[0,T]$, the operator $L_2$ has an eigenfunction
   $r$ corresponding to zero eigenvalue and the rest of the spectrum
   is contained in $(\alpha^2\lambda_1,\infty)$.

In the formulas which follow, we take $r=r(\bar{x})$ with an argument
$\bar{x}=x-\xi-2vt$. The values of $\xi$ and $\eta$ are chosen so that the
infimum in $(\ref{2.5b})$ is attained at that point.
Therefore the derivative of $d_{q}^{2}(u, \varphi)$ with respect to $\eta$
 is equal to zero. Together with (\ref{2.2}), this yields
     \begin{equation}\label{2.8}
       \begin{array}{ll}
       0&=i\int_{0}^{T}{h_xe^{-i\eta}\overline{\varphi_x}-\varphi_xe^{i\eta}\overline{h_x}+q(he^{-i\eta}\overline{\varphi}-\varphi
       e^{i\eta}\overline{h})}dx\\
       \\
       &=2Im
       \int_{0}^{T}{(-\varphi_{xx}+q\varphi)e^{i\eta}\overline{h}}dx\\
       \\
       &=\int_{0}^{T}{\left[ \left( v^2-\omega+q+|r|
       \right)rh_2+2vr'h_1 \right]}dx.
       \end{array}
     \end{equation}
 We set $h_2=\beta r(\bar{x})+\theta$,  $\;\int_0^T\theta rdx=0$.
 Substituting in (\ref{2.8}), we obtain
$$\beta ||r||^2\left[
v^2-\omega+q+{\frac{||r^{3/2}||^2}{||r||^2}}\right]+\int_{0}^{T}{[\theta
r|r|+2vr'h_1]}dx=0.$$
Using that ${\frac{||r^{3/2}||^2}{||r||^2}}>\omega$ (see estimate A1 of Appendix 1), we obtain the estimate
$$\begin{array}{ll}
        |\beta|\,||r||& \displaystyle \leq  \frac{\left|
        \int_0^T(\theta r|r|+2vr'h_1)dx\right|}
        {(v^2+q)||r||}\\[5mm]
        &\displaystyle \leq \frac{||r^2||\cdot ||\theta||+2|v|\,
        ||r'||\cdot  ||h_1||}{(v^2+q) ||r||}\\[3mm]
        & \leq m_0(||\theta||+||h_{1}||),
       \end{array}
     $$
     where $m_0=2m_1(v,\omega)/(q+\omega^2)$ and
     $$m_1(v,\omega)=\max\limits_{c\in[-\frac13\omega^3,0]}\left(
     \frac{||r^2||}{||r||}, \frac{2|v|\,||r'||}{||r||}, {\frac{2||\, |r|r'||}{||r'||}}, \, {\frac{|v|||\, |r|r-\omega r||}{||r'||}}\right).$$
     Clearly, the first and the third terms are uniformly bounded for $\omega$
     fixed. The boundedness of the second and the fourth ones follows from
     the estimates in D1 of Appendix 1.(The third and fourth terms are included
     for later use.)
      We will use below that for $v$ and $\omega$
     fixed, $m_0\rightarrow 0$ when $q\rightarrow \infty$. Further,
       $$||h_{2}||\leq |\beta|\,||r||+||\theta||\leq m_0
       (||\theta||+||h_{1}||)+||\theta||=(m_0+1)||\theta||+m_0||h_1||.$$
     Hence, we obtain
     \begin{equation}\label{2.9}
       ||\theta||^{2}\geq \frac{||h_2||^2}{2(m_0+1)^2}
       -\left( {\frac{m_0}{m_0+1}}\right)^2||h_1||^2.
       \end{equation}
   Since $L_{2}r=0$ and $\langle \theta, r\rangle=0$, then from the spectral
   properties of the operator $L_2$, it follows
     $$
       M_2=\langle L_2 h_2, h_2\rangle =\langle L_2\theta, \theta\rangle \geq \alpha^2\epsilon_1||\theta||^2.
       $$
     From here and (\ref{2.9}), one obtains
       \begin{equation}\label{2.10}
           M_2 \geq \frac{\alpha^2 \epsilon_1}{2(m_0+1)^2}||h_2||^2
           -\frac{\alpha^2 \epsilon_1m_0^2}{(m_0+1)^2} ||h_1||^2.
       \end{equation}

    \vspace{2ex}
 \vspace{1ex}
\noindent
{\bf Estimates for $M_1$.}

\vspace{0.5cm}
\noindent
 We set
     \begin{equation}\label{2.11}
       h_1=\gamma_1\psi_0(\bar{x})+\gamma_2r'(\bar{x})
       +\theta_1, \; \; r(\bar{x})=\nu \psi_0(\bar{x})+\psi,
     \end{equation}
   where
   \begin{equation}\label{2.12}
   \langle \theta_{1}, \psi_{0}\rangle= \langle \theta_{1}, r'\rangle=
   \langle \psi, \psi_{0}\rangle= \langle \psi_0,r'\rangle=
   \langle \psi, r'\rangle=0
   \end{equation}
   and $\gamma_1$, $\gamma_2$ and $\nu$ are some constants.
   By (\ref{2.12}), we have
     $$M_{1}(h_{1})=\langle L_1h_1, h_1\rangle =\gamma_1^2\lambda_0
    \langle \psi_0,\psi_0\rangle+\langle L_1\theta_1, \theta_1\rangle.$$
    Therefore, from spectral properties of the operator $L_1$ it follows
 \begin{equation}\label{2.13}
 M_1(h_1)\geq \gamma_1^2\lambda_0||\psi_0||^2+\lambda_2||\theta_1||^2.
 \end{equation}
  The fundamental difficulty in the estimate of $M_1$ is the appearance
  of the negative term $\gamma_1^2 \lambda_0 ||\psi_0||^2$. Below, we
  are going to estimate it. From the condition
    $$P(u)=\int_0^T{|h+e^{i\omega \eta}\varphi(x-\xi,t)|^2}dx=P(\varphi)$$
   we obtain
     $$||h||^2=2Re\int_0^T e^{i\omega \eta}\varphi(x-\xi,t)\overline{h} dx
     =-2\int_0^T r h_1 dx.$$
  Then using (\ref{2.11}), we have
      $$-\frac12||h||^2=\nu \gamma_1||\psi_0||^2
      +\int_0^T \psi\theta_1 dx$$
      and therefore
   \begin{equation}\label{2.14}
   \gamma_1^2||\psi_0||^2=\frac{1}{\nu^2 ||\psi_0||^2}\left(\frac12||h||^2
        +\int_0^T \psi \theta_1 dx\right)^2.
        \end{equation}
 From (\ref{2.14}), we obtain
      \begin{equation}\label{2.15}
        \gamma_1^2||\psi_0||^2\leq \frac{1}{\nu^{2}
        ||\psi_0||^2}\left({\frac{1+d}{4}}||h||^{4}
        +{\frac{d+1}{d}}||\psi||^2||\theta_{1}||^2\right),
     \end{equation}
  where $d$ is a positive constant which will be fixed later.
  Using (\ref{2.14}) and (\ref{2.13}), we derive the inequality
    \begin{equation}\label{2.16}
      M_1\geq \left(\lambda_2+\lambda_0(1+\frac{1}{d})
      {\frac{||\psi||^2}{\nu^2||\psi_0||^2}}\right)||\theta_{1}||^{2}
      +{\frac{\lambda_0(1+d)}{4\nu^2||\psi_0||^2}}||h||^4
   \end{equation}

       Below, we will denote by $C_m$, $D_m$ positive constants.
  By using (\ref{2.15}) and (\ref{2.16}), we derive the inequality
    \begin{equation}\label{2.17}
     \begin{array}{ll}
      M_1&\geq \left(\lambda_2+\lambda_0(1+\frac{1}{d})
      {\frac{||\psi||^2}{\nu^2||\psi_0||^2}}\right)||\theta_{1}||^{2}
      +{\frac{\lambda_0(1+d)}{4\nu^2||\psi_0||^2}}||h||^4 \\[2mm]
      &\geq C_1\lambda_2||\theta_1||^2-D_1||h||^4
     \end{array}
   \end{equation}
(see the estimates in point C1 of the Appendix 1).

We denote $\vartheta=h_1-\gamma_2r'(\bar{x})=\gamma_1\psi_0(\bar{x})+\theta_1$.
Then from (\ref{2.11}), (\ref{2.17}), we have
    $$\begin{array}{rl}
    ||\vartheta||^2=\gamma_1^2||\psi_0||^2+||\theta_1||^2&
 \leq \left(1+\frac{(d+1)||\psi||^2} {d\nu^2||\psi_0||^2}\right)||\theta_1||^2
       +\frac{1+d}{4\nu^2||\psi_0||^2}||h||^4\\[2mm]
    &\leq C_2||\theta_1||^2+ D_2||h||^4.\end{array}$$
    Then
      $$||\theta_1||^2\geq \frac{||\vartheta||^2}{C_2}
      -\frac{D_2||h||^4}{C_2|a|^\frac12}$$
      and hence, by (\ref{2.17}),
      \begin{equation}\label{2.18}
           M_1\geq \frac{C_1\lambda_2}{C_2}||\vartheta||^2
            -\frac{C_1\lambda_2 D_2+C_2D_1}{C_2}||h||^4.
      \end{equation}

      After differentiating (\ref{2.5b}) with respect to $\xi$ and using (\ref{2.2}), we obtain
        $$\begin{array}{ll}
  0&=2Re\int_0^T{e^{i\omega \eta}(\varphi_{xx}\overline{h}_x
  +q\varphi_x\overline{h})}dx\\[2mm]
 &=2Re\int_0^T{[(r''+2ivr'-v^2r)(h_{1x}-ih_{2x}-ivh_1-vh_2)+q(r'+ivr)(h_1-ih_2)]}dx\\[2mm]
  &=2\int_0^T{[(-\omega+2|r|+3v^2+q)r'h_1+v(v^2-3\omega+3|r|+q)rh_2]}dx.
           \end{array}
         $$
       From (\ref{2.8}), we have
 $$\int_0^T{qrh_2}dx=-\int_0^T[(v^2-\omega +|r|)rh_2+2vr'h_1]dx$$
         and replacing in the above equality, we obtain
         $$\int_0^T{[(-\omega+2|r|+v^2+q)r'h_1+v (2|r|-2\omega)rh_2]}dx=0.$$
       Substituting $h_1=\gamma_2r'(\bar{x})+\vartheta$ in the above
       equality and using the orthogonality condition
       $\langle r',\vartheta\rangle=\langle r', \gamma_1\psi_0
       +\theta_1\rangle=0$, we obtain
       $$\gamma_2||r'||^2\left(-\omega+v^2+q+\frac{2||\sqrt{r}r'||^2}
     {||r'||^2}\right)+\int_0^T{[v (2|r|-2\omega)rh_2+2|r|r'\vartheta]}dx=0.$$
   As $\frac{2||\sqrt{r}r'||^2} {||r'||^2}\geq \omega$
   (see estimate B1 from Appendix 1),

$$\begin{array}{ll}
        |\gamma_2|\,||r'||& \displaystyle\leq {\frac{\left| \int_0^T
        {[v(2|r|-2\omega)rh_2+2|r|r'\vartheta]}dx\right| }
        {(v^2+q) ||r'||}}\\[5mm]
   &\displaystyle \leq 2{\frac{|v|\,|||r|r-\omega r||\cdot||h_2||
      +2|||r|r'||\cdot ||\vartheta||}{(v^2+q)||r'||}}\\[4mm]
             &\leq m_0(||\vartheta||+||h_2||).
            \end{array}
          $$
 Hence
       $$||h_1||\leq|\gamma_2|\,||r'||
        +||\vartheta||\leq (m_0+1)||\vartheta||+m_0||h_2||,$$
        which yields
  $$||\vartheta||^{2}\geq \frac{||h_1||^2}{2(m_0+1)^2}
  -\left(\frac{m_0}{m_0+1}\right)^2||h_2||^2.$$
        Replacing in (\ref{2.18}), we finally obtain
          \begin{equation}\label{2.19}
   M_{1}\geq \frac{C_1\lambda_2}{C_2 2(m_0+1)^2}||h_1||^2
   -\frac{C_1\lambda_2m_0^2}{C_2(m_0+1)^2}||h_2||^2-{\frac{C_1\lambda_2 D_2+C_2D_1}{C_2}}|h||^4.
            \end{equation}

       \vspace{2ex}
       \noindent
      {\bf The estimate for $\Delta M$.}

\vspace{2ex}
\noindent
From (\ref{2.10}) and (\ref{2.19}), we have (fixing $q$ large and therefore $m_0$ small enough)
  $$\begin{array}{ll}
    M_1+M_2&\geq \left( {\frac{C_1\lambda_2-2C_2\alpha^2\epsilon_1
    m_0^2}{2C_2(m_0+1)^2}}\right)||h_1||^2+\left(
    {\frac{C_2\alpha^2\epsilon_1-2C_1\lambda_2
    m_0^2}{2C_2(m_0+1)^2}}\right)||h_2||^2\\
    \\
    &-{\frac{C_1\lambda_2D_2+C_2D_1}{C_2}}||h||^4
      \geq C_3 ||h||^2-D_3||h||^4.
      \end{array}
      $$
  On the other hand, estimating directly $I_{2}$ from below (for this purpose
  we use its initial formula), we obtain
       $$\begin{array}{ll}
       I_2&\geq ||h_{x}||^{2}+\int_{0}^{T}{(\omega+v^2-2|\varphi|)|h|^{2}}dx
       -2|v|\int_{0}^{T}{|h|\cdot |h_{x}|}dx\\[2mm]
       &\geq ||h_{x}||^{2}+(\omega+v^2-2\max |r|)||h||^{2}-2v^2||h||^{2}
       -{\frac{1}{2}}||h_{x}||^{2}\\[2mm]
       &={\frac{1}{2}}||h_{x}||^{2}+(\omega-v^2-2r_0)||h||^{2}.
       \end{array}$$
       Let $0<m<{\frac{1}{2}}$. We have
         $$\begin{array}{ll}
           \Delta M &=2mI_2+(1-2m)(M_1+M_2)+I_3\\
           \\
           &\geq
           m||h_x||^2+2m(\omega-v^2-2r_0)||h||^{2}+(1-2m)(C_3||h||^2-D_3||h||^4)+I_3\\
           \\
           &\geq
           m||h_x||^2+[(2m(\omega-v^2-2r_0)+(1-2m)C_3]||h||^2-(1-2m)D_3||h||^4-|I_3|.
         \end{array} $$
         We choose $m$, so that $2mq=(1-2m)C_3+2m(\omega-v^2-2r_0)$, i.e.
         $$2m={\frac{C_3}{q+C_3-\omega+v^2+2r_0}}.$$
         From the continuity of $|z|$ and $|z|e^{-2iargz}$, we have
           $$|I_3|<{\frac{mq}{2}}||h||^2$$
            From the inequality
         $$|h|^2\leq {\frac{1}{T}}\int_0^T{|h|^2}dx +2\left(\int_0^T{|h|^2}dx
         \int_0^T{|h_{x}|^2}dx \right)^\frac12$$
       we obtain
         $$|h|^2\leq {\frac{1}{T}}\int_0^T{|h|^2}dx+\sqrt{q}\int_0^T{|h|^2}dx
         +\frac{1}{\sqrt{q}}\int_0^T{|h_{x}|^2}dx.$$
         Hence for sufficiently large $q$, we obtain
       $$\max |h(x,t)|^2\leq {\frac{2}{\sqrt{q}}}d_{q}^{2}(u, \varphi )$$
    and moreover $||h||^2\leq q^{-1}d_{q}^{2}(u, \varphi )$. Consequently we
    can choose $\delta_{0}>0$, such that for $d_{q}(u, \varphi )<\delta_{0}$,
  we will have
  $[\max(4|a|^\frac12|h|+|h|^2)+(1-2m)D_3|a|^\frac12]||h||^2\leq qm$.

  Finally, we obtain that if $d_q(u, \varphi )<\delta_0$,
  then $\Delta M\geq md_q^2(u, \varphi)$.  Proposition \ref{p21} is
  completely proved.  $\Box$

\vspace{1cm}
 \noindent
{\bf Proof of Theorem \ref{t21}.} We split the proof of our main result
 into two steps. We begin with the special case
      $P(u)=P(\varphi)$. Assume that $m,q, \delta_{0}$
     have been selected according to Proposition \ref{p21}. Since
     $\Delta M$ does not depend on $t, t\in [0, \infty)$, there
     exists a constant $l$ such that $\Delta M \leq ld^{2}(u,
     \varphi)|_{t=0}$. Below, we shall assume without loss of
     generality that $l\geq 1, q\geq 1$.

     Let
       $$\varepsilon >0, \; \; \delta = \min \left( \left(
       {\frac{m}{lq}}\right){\frac{\delta_{0}}{2}}, \left(
       {\frac{m}{l}}\right)^{1/2}\varepsilon \right) $$
     and $d(u, \varphi)|_{t=0}<\delta$. Then
       $$d_{q}(u, \varphi)\leq q^{1/2}d(u,
       \varphi)|_{t=0}<{\frac{\delta_{0}}{2}} $$
       and Lemma \ref{l21}  yields that there exists a
       $t_{0}>0$ such that $d_{q}(u, \varphi)<\delta_{0}$ if $t\in [0,
       t_{0})$. Then, by virtue of Proposition \ref{p21} we have
         $$\Delta M \geq md^{2}_{q}(u, \varphi), \; \; t\in [0,
         t_{0}).$$
         Let $t_{max}$ be the largest value such that
          $$\Delta M \geq md^{2}_{q}(u, \varphi), \; \; t\in [0,
         t_{max}).$$
         We assume that $t_{max}<\infty.$ Then, for $t\in [0,
         t_{max}]$ we have
         $$d^{2}_{q}(u, \varphi)\leq {\frac{\Delta M}{m}}\leq
         {\frac{l}{m}}d^{2}(u,
         \varphi)|_{t=0}<{\frac{l}{m}}\delta^{2}\leq
         {\frac{\delta^{2}_{0}}{4}}.$$
         Applying once again Lemma \ref{l21}, we obtain that there
         exists $t_{1}>t_{max}$ such that
           $$d_{q}(u, \varphi)<\delta_{0}, \; \; t\in [0,
           t_{1}).$$
           By virtue of the proposition, this contradicts the
           assumption $t_{max}<\infty$. Consequently,
           $t_{max}=\infty$,
           $$\Delta M \geq md^{2}_{q}(u, \varphi)\geq md^{2}(u,
           \varphi), \; \; t\in [0, \infty).$$
           Therefore,
             $$d^{2}(u, \varphi)\leq {\frac{\Delta M}{m}}\leq
             {\frac{l}{m}}\delta^{2}<\varepsilon^{2}, \; \; t\in
             [0, \infty), $$
             which proves the theorem in the special case.

              Now we proceed to release the restriction
     $P(u)=||u||^{2}=||\varphi||^{2}=P(\varphi)$. We have
     $$||\varphi||=(16\alpha^3\sqrt{1-k^2+k^4}\left[
     (k^2-2+\sqrt{1-k^2+k^4})K(k)+3E(k)\right])^{1/2}.$$
     Below, we are going to apply a perturbation argument, freezing for
     a while the period $T$ and the parameters $\omega,c$ in (\ref{2.3}).
     We claim there are respective parameter values $\omega^*, c^*$, and corresponding
     $\varphi^*$, $r^*$,  $\alpha^*$, $k^*$, see (\ref{2.2}), (\ref{2.3}) and (\ref{2.5}),
     such that $\varphi^*$ has a period $T$ in $x$ and moreover,
     $||\varphi^*||=||u||$. By (\ref{2.6}), we obtain the equations
     \begin{equation}\label{ift}
     \begin{array}{l}
     \displaystyle \frac{2K(k^*)}{\alpha^*}-T=0,\\[3mm]
     \displaystyle ||r^*||^2-||u||^2=0.
     \end{array}
     \end{equation}
Moreover, one has $||\varphi^*||=||u||$ and we could use the
restricted result we established above. As $k=k^*(T,
||\varphi||)$,  $\alpha=\alpha^*(T, ||\varphi||)$, it remains to
apply the implicit function theorem to (\ref{ift}). Since the
corresponding Jacobian determinant reads
$$\left|\begin{array}{cc}
\frac{\partial}{\partial k^*}\left(\frac{2K(k^*)}{\alpha^*}\right) &
\frac{\partial}{\partial\alpha^*} \left(\frac{2K(k^*)}{\alpha^*}\right) \\[2mm]
 \frac{\partial}{\partial k^*} ||r^*||^2  &    \frac{\partial}{\partial \alpha^*} ||r^*||^2
 \end{array} \right|>0 $$
   the needed  properties are established.

   \vspace{1cm}
  \begin{figure}[h]
\centering
\includegraphics[width=5cm,height=5cm]{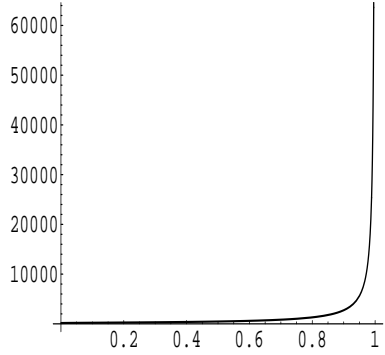}
\caption{Graph of the Jacobian determinant}
\label{fig1}
\end{figure}
   By (\ref{ift}) and our assumption, we have
     \begin{equation}\label{t21.1}
       {\frac{K(k)}{\alpha}}={\frac{K(k^*)}{\alpha^*}}={\frac{T}{2}}.
      \end{equation}
             Next, choosing
              $\eta=2(\omega^{*}-\omega)t, \; \; \xi=0$, we obtain
              inequality
              $$d^{2}(\varphi^{*}, \varphi)\leq
              (1+v^{2})||r^{*}-r||^{2}+||r^{*'}-r^{'}||^{2}.$$
              Denote for while
              $\Phi(\rho)=2\alpha^2(1-2k^2+\sqrt{1-k^2+k^4}+8k^2
              cn^2(\alpha x; k))$, where
              $k=k(\rho)$ is determined from
              $K(k)=\frac{1}{2}\rho T$. Then
              $r^{*}-r=\Phi(\alpha^*)-\Phi(\alpha)=(\alpha^*-\alpha)\Phi'(\rho)$
              with some appropriate $\rho$. Moreover,
                 $|\Phi'(\rho)|\leq C_0$ with constant
                independent of the values with $*$ accent. Hence
                $|r^{*}-r|\leq C_0|\alpha^*-\alpha|$. Similarly $|r'^{*}-r'|\leq
                C_1|\alpha^*-\alpha|$. All this, together with (\ref{t21.1})
                yields
              \begin{equation}\label{t21.2}
               d(\varphi^{*}, \varphi)\leq
              C|\alpha^*-\alpha|C={\frac{2C}{T}}|K(k^*)-K(k)|={\frac{2C}{T}}|K'(k)| |k^*-k|.
              \end{equation}

              Let $\varepsilon >0$. From the inequalities
                $$\left| ||\varphi^{*}||-||\varphi||\right|=\left|
                ||u||-||\varphi||\right|\leq d(u,
                \varphi)|_{t=0}<\delta$$
                it follows that
                 $$-||r||\delta <
                 (||\varphi||)^{-1}||\varphi^{*}||-1<||r||\delta $$
                 and, consequently,
                  $|\,||r^*||^2-||r||^2\,|<||r||^2\delta_{1}$,
                 where $\delta_{1}=(1+||r||\delta)^{2}-1$.

                 On the other hand, we have (using (\ref{t21.1})
                 again)
                   \begin{equation}\label{t21.3}
                       \begin{array}{ll}
                       |\,||r^*||^2-||r||^2\,|\\
                       \\
                       ={\frac{2^{7/2}}{T^{3/2}}}\left|
(K(k^*))^{3/2}\sqrt{1-{k^*}^2+{k^*}^4}\left[
     ({k^*}^2-2+\sqrt{1-{k^*}^2+{k^*}^4}K(k^*)+3E(k^*)\right]\right.\\
     \\
     \left. -(K(k))^{3/2}\sqrt{1-k^2+k^4}\left[
     (k^2-2+\sqrt{1-k^2+k^4}K(k)+3E(k)\right]\right|\\
     \\
     \geq C_2||k^*-k|,\end{array}
                  \end{equation}
         Thus combining (\ref{t21.2}) and (\ref{t21.3}), we get
                     $$d(u, \varphi^{*})|_{t=0}\leq d(u,
                     \varphi)|_{t=0}+d(\varphi,
                     \varphi^{*})|_{t=0}<\delta+||r||^2\widetilde{C}\delta_{1}=\delta_{0}.$$
                     We select $\delta$ sufficiently small and
                     apply the part of the theorem which has been
                     already proved,
                     $$d(u,
                     \varphi^{*})|_{t=0}<\delta_{0}\Rightarrow
                     d(u, \varphi^{*})<{\frac{\varepsilon}{2}}, \;
                     \; t\in [0, \infty).$$
                      Choosing an
                     appropriate $\delta>0$, we obtain that
                        $$d(u, \varphi)\leq d(u,
                     \varphi^{*})+d(\varphi,
                     \varphi^{*})<\varepsilon,$$
                     for all $t\in [0, \infty)$. Theorem \ref{t21} is
                     completely proved. $\Box$

%%%%%%%%%%%%%%%%%%%%%%%%%%%%%%%%%%%%%%%%%%%%%%%%%%%%%%%%%%%%%%%%%%%%%%%%

\section{Existence and stability of periodic traveling waves for the
cubic Schr\"odinger equation}

 Consider the cubic nonlinear Schr\"odinger equation
   \begin{equation}\label{3.1}
    iu_{t}+u_{xx}+|u|^{2}u=0,
   \end{equation}
  where $u=u(x,t)$ is a complex-valued function of $(x,t)\in\R^2$.

 Equation (\ref{3.1}) possesses the following family of
 traveling-wave solutions
    \begin{equation}\label{3.1a}
    \varphi(x,t)=e^{i(vx+(\omega-v^{2})t)}r(x-2vt),
    \end{equation}
  where $\omega$ and $v$ are real parameters and the real-valued function
  $r(x)$ satisfies equation
    \begin{equation}\label{3.2}
      r''-\omega r+r^{3}=0.
    \end{equation}
    Integrating once again, we obtain
 \begin{equation}\label{3.3}
 r'^2-\omega r^2+\frac12 r^4=c
 \end{equation}
  and $r$ is a periodic function
provided that the energy level set $H(x,y)=c$ of the Hamiltonian
system $dH=0$, $$H(x,y)=y^2-\omega x^2+\frac12x^4,$$ contains an
oval (a simple closed real curve free of critical points). The
level set $H(x,y)=c$ contains two periodic trajectories if
$\omega>0$, $c\in(-\frac12\omega^2,0)$ and a unique periodic
trajectory if $\omega\in\R$, $c>0$. Under these conditions, the
solution of (\ref{3.2}) is determined by $H(r,r')=c$ and $r$ is
periodic of period $T=T(\omega,c)$.

Below, we are going to consider the case $c<0$. Let us denote by
$r_0>r_1>0$ the positive roots of ${\frac{1}{2}}r^4-\omega
r^2-c=0$. Then, up to a translation, we obtain the respective
explicit formulas
\begin{equation}\label{3.4}
r(z)=\mp r_0 dn(\alpha z; k),\quad k^2=\frac{r_0^2-r_1^2}{r_0^2}
=\frac{-2\omega+2r_0^2}{r_0^2}, \quad
\alpha={\frac{r_0}{\sqrt{2}}}, \quad T=\frac{2K(k)}{\alpha}.
\end{equation}
Recall that $K(k)$ and $E(k)$ are, as usual, the complete
elliptic integrals of the first and second kind in a Legendre
form. By (\ref{3.4}), one also obtains $\omega=(2-k^2)\alpha^2$
and, finally,
\begin{equation}\label{3.5}
T=\frac{2\sqrt{2-k^2}K(k)}{\sqrt{\omega}}, \quad k\in(0,1), \quad
T\in I=\left(\frac{2\pi}{\sqrt{\omega}},\infty\right).
\end{equation}

 \vspace{1ex}
\noindent We take $\omega>0$, an energy level
$c\in(-\frac12\omega^2,0)$ and let $T$ be the (minimal) period of
$r(x)$. Assume that $v\neq 0$ is chosen to satisfy $vT/2\pi\in\Z$.
Then there are two periodic solutions $r$ of (\ref{3.2}): left
$(r<0)$  and right $(r>0)$ and the corresponding traveling-wave
solution $\varphi(x,t)$ is periodic with respect to $x$ of period
$T$.

\vspace{2ex} \noindent \begin{lem}\label{l31}
 For any $\omega>0$ and
$T\in I$, there is a constant $c=c(\omega)$ such that the periodic
traveling-wave solution $(\ref{3.4})$ determined by
$H(r,r')=c(\omega)$ has a period $T$. The function $c(\omega)$ is
differentiable.
\end{lem}

\noindent
{\bf Proof.} See \cite{HIK2}, Lemma 3.1.

\vspace{1ex}
Equation (\ref{3.1}) has the following conservation laws
  $$
     Q(u)=i\int_{0}^{T}{\overline{u}_{x}u}dx,\quad
     P(u)=\int_{0}^{T}{|u|^{2}}dx,\quad
     \tilde{E}(u)=\int_{0}^{T}{(|u_{x}|^{2}-{\frac{|u|^{4}}{2}})}dx.
  $$
  Let us consider the functional
    $$\tilde{M}(u)=\tilde{E}(u)+(\omega +v^{2})P(u)-2vQ(u).$$

\begin{thm}\label{t31}   Let $\varphi$ be given by $(\ref{3.1a})$, with $r\neq 0$.
         For each $\varepsilon>0$ there exists $\delta>0$ such that if
         $u(x,t)$ is a solution of $(\ref{3.1})$ and
         $d(u, \varphi )_{|t=0}<\delta$, then $d(u, \varphi)<\varepsilon$
         $\forall t\in [0,\infty)$.
       \end{thm}
The crucial step in the proof will be to verify the following
statement.
 \vspace{1ex}
\noindent

\begin{prop}\label{p31}
  There exist positive constants $m, q, \delta_{0}$ such that if $u$ is
a periodic solution of $(\ref{3.1})$,  $u(x,t)=u(x+T,t)$,
$P(u)=P(\varphi)$ and $d_{q}(u, \varphi)<\delta_{0}$, then
    \begin{equation}\label{3.8}
    \tilde{M}(u)-\tilde{M}(\varphi)\geq  m d_{q}^{2}(u, \varphi).
    \end{equation}
 \end{prop}

\noindent
{\bf Proof.} In order to estimate $\Delta
\tilde{M}=\tilde{M}(u)-\tilde{M}(\varphi)$, we set
   $$u(x,t)=e^{i\eta}\varphi(x-\xi, t)+h(x,t),$$
   $$\zeta =v(x-\xi)+(\omega+v^2)t+\eta ,$$
   and integrating by parts in the terms containing $h_{x}$ and
   $\overline{h}_{x}$, we obtain
     $$\begin{array}{ll}
      \Delta \tilde{M} &=\tilde{M}(u)-\tilde{M}(\varphi)\\
      \\
      &=2Re\int_0^T{e^{i\eta}[-\varphi_{xx}
      +(\omega+v^2-|\varphi|^2)\varphi+2iv\varphi_x]\overline{h}}dx\\
      \\
      &+\int_0^T{[|h_{x}|^2+(\omega+v^2-2|\varphi|^2)|h|^2
      -|\varphi|^2Re(e^{-2i\zeta}h^2)-2ivh\overline{h}_x]}dx\\
      \\
      &-{\frac12}\int_0^T{|h|^2(4Re(e^{i\eta}\varphi\overline{h})+|h|^2)}dx\\
      \\
      &=I_{1}+I_{2}+I_{3}.
        \end{array}
     $$
   Using that $r(x)$ satisfies the equation (\ref{3.2}) we obtain that
   $I_{1}=0$.

   Let
     $$h(x,t)=(h_{1}+ih_{2})e^{i\zeta},$$
   where $h_{1}$ and $h_{2}$ are real periodic functions with period $T$.
   Using (\ref{2.7}), for $I_2$ we obtain

       $$
          I_{2}=\int_0^T{[h_{1x}^2+(\omega-3r^2) h_1^2 ]}dx
          +\int_0^T{[h_{2x}^2 +(\omega -r^2)h_2^2]}dx=\tilde{M}_1+\tilde{M}_2
       $$
       Introduce in $L^2[0,T]$ the self-adjoint operators $L_1$ and $L_2$
       generated by the differential expressions
       \begin{equation}\label{oper}
        \begin{array}{ll}
           L_{1}=-\partial_{x}^{2}+(\omega-3r^{2}), \\[2mm]
           L_{2}=-\partial_{x}^{2}+(\omega -r^{2}),
          \end{array}
          \end{equation}
         with periodic boundary conditions in $[0,T]$.

         \vspace{2ex}
     We use now (\ref{3.4}) and (\ref{3.5}) to rewrite operators
     $L_1$, $L_2$ in more
     appropriate form. From the expression for $r(x)$ from (\ref{3.4})
     and the relations
     between elliptic functions $sn(x)$, $cn(x)$ and $dn(x)$, we obtain
       $$L_1=\alpha^{2}[ -\partial_{y}^{2}+6k^{2} sn^{2}(y)-4-k^2] $$
     where $y=\alpha x$.

     It is well-known that the first five eigenvalues of
     $\Lambda_1=-\partial_{y}^{2}+6k^{2}sn^{2}(y, k)$,
     with periodic boundary conditions on $[0, 4K(k)]$, where
     $K(k)$ is the complete elliptic integral of the first kind, are
     simple. These eigenvalues and corresponding eigenfunctions are:
      $$\begin{array}{ll}
         \nu_{0}=2+2k^2-2\sqrt{1-k^2+k^4},
         & \phi_{0}(y)=1-(1+k^2-\sqrt{1-k^{2}
         +k^{4}})sn^{2}(y, k),\\[1mm]
         \nu_{1}=1+k^{2}, & \phi_{1}(y)=cn(y, k)dn(y, k)
         =sn'(y, k),\\[1mm]
         \nu_{2}=1+4k^{2}, & \phi_{2}(y)=sn(y, k)dn(y, k)
         =-cn'(y, k),\\[1mm]
         \nu_{3}=4+k^{2}, & \phi_{3}(y)=sn(y, k)cn(y, k)
         =-k^{-2}dn'(y, k),\\[1mm]
         \nu_{4}=2+2k^{2}+2\sqrt{1-k^{2}+k^{4}},
         & \phi_{4}(y)=1-(1+k^{2}+\sqrt{1-k^{2}
         +k^{4}})sn^{2}(y, k).
        \end{array}
      $$

      It follows that the first three eigenvalues of the operator
      $L_1$, equipped with periodic boundary condition on $[0,2K(k)]$
      (that is, in the case of left and right family),
      are simple and $\lambda_0=\alpha^2(\nu_0-\nu_3)<0, \;
      \lambda_1=\alpha^2(\nu_3-\nu_3)=0, \;
      \lambda_{2}=\alpha^2(\nu_4-\nu_3)>0$.
The corresponding eigenfunctions are $\psi_0=\phi_0(\alpha x),
\psi_1=r'(x), \psi_2=\phi_4(\alpha x)$.

\vspace{2ex} Similarly, for the operator $L_2$ we have
$$L_2=\alpha^2[-\partial_y^2+2k^2sn^2(y, k)-k^2]$$
in the case of left and right family. The spectrum of
 $\Lambda_2=-\partial_y^2+2k^{2}sn^{2}(y, k)$ is formed
 by bands $[k^{2}, 1]\cup [1+k^{2}, +\infty)$. The
 first three eigenvalues and the corresponding eigenfunctions with
 periodic boundary conditions on $[0, 4K(k)]$ are simple and
       $$\begin{array}{ll}
          \epsilon_0=k^2, & \theta_0(y)=dn(y, k),\\[1mm]
          \epsilon_1=1, & \theta_1(y)=cn(y, k),\\[1mm]
          \epsilon_2=1+k^2, & \theta_2(y)=sn(y, k).
        \end{array}
      $$

From (\ref{3.3}) it follows that zero is an eigenvalue of  $L_2$
and it is the first eigenvalue in the case of left and right
family, with corresponding eigenfunction $r(x)$.

\vspace{2ex}
\noindent
{\bf Estimates for $\tilde{M}_2$.}

\vspace{1ex}
\noindent
As in Section 2, we use below $r=r(\bar{x})$ with an argument
$\bar{x}=x-\xi-2vt$.
From the explanations above, we know that when considered in $[0,T]$,
the operator $L_2$ has an eigenfunction
   $r$ corresponding to zero eigenvalue and the rest of the spectrum
   is contained in $(\alpha^2,\infty)$.

     The derivative of $d_{q}^{2}(u, \varphi)$ with respect to $\eta$
     at the point where the minimum is attained is equal to zero.
     Together with (\ref{3.2}), this yields
     \begin{equation}\label{3.9}
       \int_{0}^{T}{\left[ \left( v^2-\omega+q+r^2
       \right)rh_2+2vr'h_1 \right]}dx=0
     \end{equation}
     We set $h_2=\beta r(\bar{x})+\theta$,  $\;\int_0^T\theta rdx=0$.
Substituting in (\ref{3.9}), we obtain
$$\beta ||r||^2\left[
v^2-\omega+q+{\frac{||r^2||^2}{||r||^2}}\right]+\int_{0}^{T}{[\theta
r^3+2vr'h_1]}dx=0.$$
Using that $\frac{2||r^2||^2}{||r||^2} \geq w$ (see estimate A2 of the Appendix2),
    we obtain the estimate
      $$\begin{array}{ll}
        |\beta|\,||r||& \displaystyle \leq  \frac{\left|
        \int_0^T(\theta r^3+2v r' h_1)dx\right|}
        {(q+v^2)||r||}\\[5mm]
        &\displaystyle \leq \frac{||r^3||\cdot ||\theta||+2|v|\,
        ||r'||\cdot  ||h_1||}{(q+v^2) ||r||}\\[3mm]
        & \leq m_0(||\theta||+||h_{1}||),
       \end{array}
     $$
     where $m_0=2m_1(v,\omega)/(q+\omega^2)$ and
     $$m_1(v,\omega)=\max\limits_{c\in[-\frac12\omega^2,0]}\left(
     \frac{||r^3||}{||r||}, \frac{2|v|\,||r'||}{||r||}, \, {\frac{2|v|\, ||r^3-\omega r||}{||r'||}}, \, {\frac{3||r^2r'||}{||r'||}}\right)$$
     (the third and fourth item are included for later use). It is obvious
     that the first and the last fractions are bounded. For the second and the
     third ones, see
     estimates D2 in Appendix 2. We will use below that for $v$ and $\omega$
     fixed, $m_0\rightarrow 0$ when $q\rightarrow \infty$. Further,
       $$||h_{2}||\leq |\beta|\,||r||+||\theta||\leq m_0
       (||\theta||+||h_{1}||)+||\theta||=(m_0+1)||\theta||+m_0||h_1||.$$
     Hence, we obtain
     \begin{equation}\label{3.10}
       ||\theta||^{2}\geq \frac{||h_2||^2}{2(m_0+1)^2}
       -\left( {\frac{m_0}{m_0+1}}\right)^2||h_1||^2.
       \end{equation}
   Since $L_{2}r=0$ and $\langle \theta, r\rangle=0$, then from the spectral
   properties of the operator $L_2$, it follows
     $$
       M_2=\langle L_2 h_2, h_2\rangle =\langle L_2\theta, \theta\rangle\geq
       \alpha^2\langle \theta, \theta\rangle \geq \frac{\omega}{2}||\theta||^2.
       $$
     From here and (\ref{3.10}), one obtains
       \begin{equation}\label{3.10a}
           M_2 \geq \frac{\omega}{4(m_0+1)^2}||h_2||^2
           -\frac{\omega m_0^2}{2(m_0+1)^2} ||h_1||^2.
       \end{equation}
 \vspace{1ex}
\noindent

 \vspace{1ex}
  \noindent
{\bf Estimates for $\tilde{M}_1$.}

\vspace{1ex}
\noindent
We set
     \begin{equation}\label{3.11}
       h_1=\gamma_1\psi_0(\bar{x})+\gamma_2r'(\bar{x})
       +\theta_1, \; \; r(\bar{x})=\nu \psi_0(\bar{x})+\psi,
     \end{equation}
   where
   \begin{equation}\label{3.12}
   \langle \theta_{1}, \psi_{0}\rangle= \langle \theta_{1}, r'\rangle=
   \langle \psi, \psi_{0}\rangle= \langle \psi_0,r'\rangle=
   \langle \psi, r'\rangle=0
   \end{equation}
   and $\gamma_1$, $\gamma_2$ and $\nu$ are some constants.
As the same line as for $M_1$, we obtain
  \begin{equation}\label{3.13}
    \gamma_1^2||\psi_0||^2=\frac{1}{\nu^2 ||\psi_0||^2}\left(\frac12||h||^2
        +\int_0^T \psi \theta_1 dx\right)^2.
        \end{equation}
        and
 \begin{equation}\label{3.14}
    \begin{array}{ll}
      {\tilde{M}_1} &\geq \left(\lambda_2+\lambda_0(1+\frac{1}{d})
      {\frac{||\psi||^2}{\nu^2||\psi_0||^2}}\right)||\theta_{1}||^{2}
      +{\frac{\lambda_0(1+d)}{4\nu^2||\psi_0||^2}}||h||^4\\
      \\
      &\geq C_1\lambda_2||\theta_1||^2-D_1w||h||^4.
      \end{array}
      \end{equation}
(see the estimates in point C2 of the Appendix2).

We denote
$\vartheta=h_1-\gamma_2r'(\bar{x})=\gamma_1\psi_0(\bar{x})+\theta_1$.
Then from (\ref{3.13}), (\ref{3.14}) and (since
$\lambda_2=-\omega+\sqrt{4\omega^2+6c}$ and
$\lambda_2=-\omega-\sqrt{4\omega^2+6c}$ the inequalities
$\lambda_2\leq\frac13|\lambda_0|\leq w$, we have
    $$\begin{array}{rl}
    ||\vartheta||^2=\gamma_1^2||\psi_0||^2+||\theta_1||^2&
 \leq \left(1+\frac{(d+1)||\psi||^2} {d\nu^2||\psi_0||^2}\right)||\theta_1||^2
       +\frac{1+d}{4\nu^2||\psi_0||^2}||h||^4\\[2mm]
    &\leq C_2||\theta_1||^2+ D_2\omega^{-\frac12}||h||^4.\end{array}$$
    Then
      $$||\theta_1||^2\geq \frac{||\vartheta||^2}{C_2}
      -\frac{D_2||h||^4}{C_2\omega^\frac12}$$
      and hence, by (\ref{3.14}) and $\lambda_2\leq w$,
      \begin{equation}\label{3.15}
        \begin{array}{rl}
           M_1&\displaystyle\geq \frac{C_1\lambda_2}{C_2}||\vartheta||^2
            -\frac{C_1D_2+C_2D_1}{C_2}\omega^\frac12||h||^4\\[3mm]
            &= C_3\lambda_2||\vartheta||^2-D_3\omega^\frac12||h||^{4}.
        \end{array}
      \end{equation}

      After differentiating (\ref{2.5b}) with respect to $\xi$ and using (\ref{3.2}), we obtain
        $$\begin{array}{ll}
  0&=2Re\int_0^T{e^{i\eta}(\varphi_{xx}\overline{h}_x
  +q\varphi_x\overline{h})}dx\\[2mm]
  &=2\int_0^T{[(-\omega+3v^2+3r^2+q)r'h_1+v(-3\omega+v^2 +3r^2+q)rh_2]}dx.
           \end{array}
         $$
       From (\ref{3.9}), we have
 $$\int_0^T{qrh_2}dx=-\int_0^T[2v r'h_1+(-\omega+v^2+r^2)rh_2]dx$$
         and replacing in the above equality, we obtain
         $$\int_0^T{[(-\omega+v^2+3r^2+q)r'h_1+v (-2\omega+2r^2)rh_2]}dx=0.$$
       Substituting $h_1=\gamma_2r'(\bar{x})+\vartheta$ in the above
       equality and using the orthogonality condition
       $\langle r',\vartheta\rangle=\langle r', \gamma_1\psi_0
       +\theta_1\rangle=0$, we obtain
       $$\gamma_2||r'||^2\left(-\omega+v^2+q+\frac{3||rr'||^2}
     {||r'||^2}\right)+\int_0^T{[2v (-\omega+r^2)rh_2+3r^2r'\vartheta]}dx=0.$$
         Using that ${\frac{3||rr'||^2}{||r'||^2}}\geq \omega$ (see the estimates in point  B2 of the Appendix2),
         we further have
           $$\begin{array}{ll}
        |\gamma_2|\,||r'||& \displaystyle\leq {\frac{\left| \int_0^T
        {[2v (-\omega+r^2)rh_2+3r^2r'\vartheta]}dx\right| }
        {(v^2+q) ||r'||}}\\[5mm]
   &\displaystyle \leq {\frac{2|v|\,||-\omega r+r^3||\cdot||h_2||
      +3||r^2r'||\cdot ||\vartheta||}{(v^2+q)||r'||}}\\[4mm]
             &\leq m_0(||\vartheta||+||h_2||).
            \end{array}
          $$
 Hence
       $$||h_1||\leq|\gamma_2|\,||r'||
        +||\vartheta||\leq (m_0+1)||\vartheta||+m_0||h_2||,$$
        which yields
  $$||\vartheta||^{2}\geq \frac{||h_1||^2}{2(m_0+1)^2}
  -\left(\frac{m_0}{m_0+1}\right)^2||h_2||^2.$$
        Replacing in (\ref{3.15}), we finally obtain
          \begin{equation}\label{3.16}
   M_{1}\geq \frac{C_3\lambda_2}{2(m_0+1)^2}||h_1||^2
   -\frac{C_3\lambda_2m_0^2}{(m_0+1)^2}||h_2||^2-D_3\omega^\frac12|h||^4.
            \end{equation}

\vspace{2ex}
\noindent
 {\bf  The estimate for $\Delta M$.}
%%%%%%%%%%%%%%%%%%%%%%%%%%%%%%%%%%%%%%%%%%%%%%%%%%%%%%%%%%%%%%%%%%%%%%%%%%%%%

\vspace{2ex}
\noindent
 From (\ref{3.10a}) and (\ref{3.16}), one obtains
    $$ M_1+M_2 \geq \frac{C_3\lambda_2-\omega m_0^2}{2(m_0+1)^2} ||h_1||^2
  +\frac{\omega-4C_3\lambda_2m_0^2}{4(m_0+1)^2}||h_2||^2-D_3\omega^\frac12||h||^4.$$
  We now fix $q$ so that $\omega m_0^2\leq\frac12C_3\lambda_2$ and assuming that
  $C_3\leq\frac12$ (which is no loss of generality), one has also
  $4C_3\lambda_2m_0^2\leq\frac12 \omega$. Therefore we come to 
     $$M_1+M_2\geq C_4\lambda_2(||h_1||^2+||h_{2}||^2)-D_3\omega^\frac12||h||^4=
   C_4\lambda_2||h||^2-D_3\omega^\frac12||h||^4$$
   where $C_4$ and $D_3$ are absolute positive constants independent on the parameters
   of the system.

  On the other hand, estimating directly $I_{2}$ from below (for this purpose
  we use its initial formula), we have
       $$\begin{array}{ll}
       I_2&\geq ||h_{x}||^{2}+\int_{0}^{T}{(\omega+v^2-2r^{2})|h|^{2}}dx
       -2|v|\int_{0}^{T}{|h|\cdot |h_{x}|}dx
       -\int_{0}^{T}{r^{2}|h|^{2}}dx\\[2mm]
       &\geq ||h_{x}||^{2}+(\omega+v^2-2r_0^2)||h||^{2}-2v^2||h||^{2}
       -{\frac{1}{2}}||h_{x}||^{2}-r_0^2||h||^{2}\\[2mm]
       &={\frac{1}{2}}||h_{x}||^{2}-(v^2+5\omega)||h||^{2}.
       \end{array}
       $$
       Similarly, $|I_{3}|\leq \max(4|2\omega|^\frac12|h|+|h|^2)||h||^2$.
       Let $0<m<\frac12$. We obtain
         $$\begin{array}{ll}
             \Delta M & =2mI_2+(1-2m)(M_1+M_2)+I_3\\[1mm]
             &\geq m||h_{x}||^2-2m(v^2+5\omega)||h||^2
             +(1-2m)(C_4\lambda_2||h||^2
             -D_3\omega^\frac12||h||^4)\\[1mm]
             &-\max (4|2\omega|^\frac12|h|+|h|^2)||h||^2\\[1mm]
             &=m||h_{x}||^2+\left[-2m(v^2+5\omega)+
             (1-2m)C_4\lambda_2\right]||h||^2\\[1mm]
             &-[\max(4|2\omega|^\frac12|h|+|h|^2)+(1-2m)D_3\omega^\frac12||h||^2]||h||^2.
             \end{array}
        $$
        We choose $m$, so that
          $$2qm=(1-2m)C_4\lambda_2-2m(v^2+5\omega),\;\;\mbox{\rm i.e.}\;\;
       2m=\frac{C_4\lambda_2}{q+v^2+5\omega+C_4\lambda_2}<1.$$
       From the inequality
         $$|h|^2\leq {\frac{1}{T}}\int_0^T{|h|^2}dx +2\left(\int_0^T{|h|^2}dx
         \int_0^T{|h_{x}|^2}dx \right)^\frac12$$
       we obtain
         $$|h|^2\leq {\frac{1}{T}}\int_0^T{|h|^2}dx+\sqrt{q}\int_0^T{|h|^2}dx
         +\frac{1}{\sqrt{q}}\int_0^T{|h_{x}|^2}dx.$$
         Hence for sufficiently large $q$, we obtain
       $$\max |h(x,t)|^2\leq {\frac{2}{\sqrt{q}}}d_{q}^{2}(u, \varphi )$$
    and moreover $||h||^2\leq q^{-1}d_{q}^{2}(u, \varphi )$. Consequently we
    can choose $\delta_{0}>0$, such that for $d_{q}(u, \varphi )<\delta_{0}$,
  we will have
  $[\max(4|2\omega|^\frac12|h|+|h|^2)+(1-2m)D_3\omega^\frac12]||h||^2\leq qm$.

  Finally, we obtain that if $d_q(u, \varphi )<\delta_0$,
  then $\Delta M\geq md_q^2(u, \varphi)$.  Proposition \ref{p31} is
  completely proved.  $\Box$

\vspace{2ex} \noindent {\bf Proof of Theorem \ref{t31}.} The proof
of the theorem in the case
$P(u)=||u||^{2}=||\varphi||^{2}=P(\varphi)$ is the same as  in
Theorem \ref{t21}.
  If
     $P(u)\neq P(\varphi)$, we proceed similarly as in Theorem \ref{t21}. We have
     $||\varphi||=(2\sqrt{2}r_0E(k))^{1/2}$, where $r_0$ is given by (\ref{3.4}).

     We claim there are respective parameter values $\omega^*, c^*$, and corresponding
     $\varphi^*$, $r^*$,  $r_0^*$, $k^*$, see (\ref{3.4}) and (\ref{3.5}),
     such that $\varphi^*$ has a period $T$ in $x$ and moreover,
     $2\sqrt{2}r_0^*E(k^*)=||u||^2$. By (\ref{3.4}), we obtain the equations
     \begin{equation}\label{ift1}
     \begin{array}{l}
     \displaystyle \frac{2\sqrt{2}K(k^*)}{r_0^*}-T=0,\\
     \displaystyle 2\sqrt{2}r_0^*E(k^*)-||u||^2=0.
     \end{array}
     \end{equation}
If (\ref{ift1}) has a solution $k^*=k^*(T, ||u||)$,
$r_0^*=r_0^*(T, ||u||)$, then the parameter values we need are
given by
$$2w^*=(2-{k^*}^2){r_0^*}^2,\quad 2c^*=({k^*}^2-1){r_0^*}^4.$$
Moreover, one has $||\varphi^*||=||u||$ and we could use the
restricted result we established above. As $k=k^*(T,
||\varphi||)$,  $r_0=r_0^*(T, ||\varphi||)$, it remains to apply
the implicit function theorem to (\ref{ift1}). Since the
corresponding Jacobian determinant reads
$$\left|\begin{array}{cc}\frac{2\sqrt{2}K'(k^*)}{r_0^*} & -\frac{2\sqrt{2}K(k^*)}{{r_0^*}^2} \\
   2\sqrt{2}r_0^*E'(k^*)  &
   2\sqrt{2}E(k^*)\end{array}\right|=\frac{8}{r_0^*}(KE)'>0
   $$
   the existence of $\omega^*$ and $c^*$ with the
   needed  properties is established.

By (\ref{ift1}) and our assumption, we have

  \begin{equation}\label{3.17}
    \frac{2\sqrt{2}K(k^*)}{r_0^*}=\frac{2\sqrt{2}K(k)}{r_0}=T.
  \end{equation}

             Choosing
              $\eta=2(\omega^{*}-\omega)t, \; \; \xi=0$, we obtain
              the inequality
              $$d^{2}(\varphi^{*}, \varphi)\leq
              (1+v^2)||r^{*}-r||^{2}+||r^{*'}-r^{'}||^{2}$$
  For $\Phi(\rho)=\rho dn (z\rho, k(\rho))$, where $k=k(\rho)$ is
  determined from $K(k)={\frac{1}{2}}\rho T$, we get
  $r^*-r=\Phi(r_0^*)-\Phi(r_0)=(r_0^*-r_0)\Phi'(\rho)$ with some
  appropriate $\rho$. Moreover, $|\Phi'(\rho)|\leq C_0$. Hence
  $|r^*-r|\leq C_0|r_0^*-r_0|$,  $|r^{*'}-r'|\leq C_1 |r_0^*-r_0|$,
  and by (\ref{3.17})
    $$d(\varphi^*,\varphi)\leq |r_0^*-r_0|\leq \frac{2\sqrt{2}C}{T}|K(k^*)-K(k)|\leq \frac{2\sqrt{2}C}{T}|K'(k)| |k^*-k|.$$
    From the inequalities
                $$\left| ||\varphi^{*}||-||\varphi||\right|=\left|
                ||u||-||\varphi||\right|\leq d(u,
                \varphi)|_{t=0}<\delta$$
               it follows that
                 $$-\left( 2\sqrt{2}r_{0}E \right)^{-1/2}\delta <
                 (||\varphi||)^{-1}||\varphi^{*}||-1<\left( 2\sqrt{2}r_{0}E
                 \right)^{-1/2}\delta $$
                 and, therefore,
                 $1-\delta_{1}<{\frac{r_{0}^{*}E(\kappa^{*})|}{r_{0}E}}<1+\delta_{1}$,
                 i.e. $|r_{0}^{*}E(\kappa^{*})-r_{0}E|<r_{0}E\delta_{1}$,
                 where $\delta_{1}=(1+(2r_{0}E)^{-1/2}\delta)^{2}-1$.
                 On the other hand
                   $$|r_0^*E(k^*)-r_0E(k)|={\frac{2\sqrt{2}}{T}}|K(k^*)E(k^*)-K(k)E(k)|={\frac{2\sqrt{2}}{T}}|(KE)'(k)|
                   |k^*-k|\geq C_2|k^*-k|,$$
                   with appropriate $C_2>0$. Thus
                     $$d(u, \varphi^{*})|_{t=0}\leq d(u,
                     \varphi)|_{t=0}+d(\varphi,
                     \varphi^{*})|_{t=0}<\delta+r_{0}E\widetilde{C}\delta_{1}=\delta_{0}.$$
            The rest of the proof is the same as in Theorem
            \ref{t21}.
              $\Box$

\vspace{2ex}
  \section{Appendix 1}
%%%%%%%%%%%%%%%%%%%%%%%%%%%%%%%%%%%%%%%%%%%%%%%%%%%%%%%%%%%%%%%%%%%%%%%%%%
Below, we provide some estimates needed in our proofs concerning the
quadratic Schr\"odinger equation. Without loss of generality, we will
assume that $r>0$. The case of negative $r$ is dealt with by changing
$r \to -r$ in all equations from Section 1.

For $n\in\Z$ and $c\in(-\frac13\omega^3,0)$, consider the
line integrals $I_n(c)$ and their  derivatives $I'_n(c)$ given by
\begin{equation}\label{in1}
I_n(c)=\oint_{H=c}x^nydx,\qquad I_n'(c)=\oint_{H=c}\frac{x^ndx}{2y}
\end{equation}
where the integration is along the
right oval contained in the level set $\{H=c\}$ and $H(x,y)=y^2-\omega x^2
+\frac23x^3$. These integrals would be useful because
\begin{equation}\label{red1}
\int_0^Tr^n(t)dt=2\int_0^{\frac12T}r^n(t)dt
=2\int_{r_1}^{r_0}\frac{x^ndx}{\sqrt{c+\omega x^2-\frac23x^3}}
=\oint_{H=c}\frac{x^ndx}{y}=2I_n'(c).
\end{equation}
(we applied a change of the variable $r(t)=x$ in the integral
and used equation (\ref{2.3})).
The properties of $I_n$ are well known, see e.g. \cite{HIK1}
for a similar treatment. Below, we list some facts we are going to use.

\vspace{2ex}
\noindent
{\bf Lemma.}  (i)  {\it The following identity holds:
$$3nc I_{n-1}+3\omega(n+3)I_{n+1}-(2n+9)I_{n+2}=0,\quad n\in\Z$$
which implies}
\begin{equation}\label{recur1}
\textstyle
I_2=\omega I_1,\quad
I_3=\frac{3}{11}cI_0+\frac{12}{11}\omega^2I_1,\quad
I_4=\frac{45}{143}c\omega I_0+(\frac{6}{13}c+\frac{180}{143}\omega^3)I_1,
\end{equation}

\vspace{1ex}
\noindent
(ii) {\it The integrals $I_0$ and $I_1$ satisfy the system}
$$\begin{array}{l}
6cI'_0+2\omega^2I'_1=5I_0,\\
6\omega cI'_0+(30c+12\omega^3)I'_1=35I_1.\end{array}$$

\vspace{1ex}
\noindent
(iii) {\it The ratio $R(c)=I_1'(c)/I_0'(c)$ satisfies the Riccati equation
and related system
\begin{equation}\label{ricc1}
6c(3c+\omega^3)R'(c)=\omega^2R^2(c)+6cR(c)-3\omega c,\qquad
\begin{array}{l}\dot{c}=6c(3c+\omega^3),\\
\dot{R}=\omega^2R^2+6cR-3\omega c,\end{array}\end{equation}
which imply estimates}
\begin{equation}\label{est1}
-\frac{3c}{\omega^2}\leq R(c)\leq\frac{5\omega}{6}-\frac{c}{2\omega^2}.
\end{equation}

\vspace{2ex}
\noindent
The equations in (i)--(iii) are derived in a standard way, see \cite{HIK1}
for more details. The estimates (\ref{est1}) follow from the fact that,
in the $(c,R)$-plane,
the graph of $R(c)$ coincides with the concave separatrix trajectory of
the system (\ref{ricc1}) contained in the triangle with vertices $(0,0)$,
$(-\frac13\omega^3,\omega)$ and $(0,\frac{5\omega}{6})$ and connecting
the first two of them.

After this preparation, we turn to prove the estimates we used in the
preceding Section 2.

\vspace{2ex}
\noindent
{\bf A1. The estimate for } $A_1=\frac{||r^{3/2}||^2}{||r||^2}$.
By (\ref{red1}), (\ref{recur1}) and the first inequality in (\ref{est1}),
we have
$$A_1=\frac{\int_0^Tr^3dt}{\int_0^Tr^2dt}=\frac{I_3'}{I_2'}=
\frac35\frac{cI_0'+2\omega^2I_1'}{\omega I_1'}=\frac{3c}{5\omega R}
+\frac{6\omega}{5}\geq\omega.$$

\vspace{2ex}
\noindent
{\bf B1. The estimate for} $B_1=\frac{2||r^{1/2} r'||^2}{||r'||^2}$.
By (\ref{2.3}) and (\ref{recur1}), we have as above
$$B_1=\frac{2\int_0^T r(c+\omega r^2-\frac23 r^3)dt}{\int_0^T (c+\omega r^2-\frac23 r^3)dt}
=\frac{2(cI_1'+\omega I_3'-\frac23 I_4')}{cI_0'+\omega I_2'-\frac23I_3'}$$
$$=\frac67\frac{c\omega I'_0+(5c+2\omega^3)I'_1}{3cI'_0+\omega^2I'_1}
=\frac67\frac{c\omega+(5c+2\omega^3)R}{3c+\omega^2R}\geq \frac{12}{7}\omega.$$
To obtain the last inequality, we used both estimates in (\ref{est1}).

\vspace{2ex}
\noindent
 {\bf C1. The estimate for}
$C_1=\lambda_2+\lambda_0(1+\frac{1}{d})\frac{||\psi||^2}{\nu^2||\psi_0||^2}$.
By (\ref{2.11}) and (\ref{2.12}) we have
$$||\psi||^2=||r||^2-\nu^2||\psi_0||^2,\;\;\mbox{\rm where}\;\;
\nu=\frac{\langle r,\psi_0\rangle}{||\psi_0||^2}.$$
Therefore
$$C_1=\lambda_2+\lambda_0\left(1+\frac{1}{d}\right)
\left(\frac{||r||^2||\psi_0||^2}{\langle r,\psi_0\rangle^2}-1\right).$$

Below, we need to use the following well-known equalities (see e.g.
\cite{ByFr}) which are written out here for reader's convenience:

\vspace{2ex}
(i) $\quad \int_0^Kdn^2xdx=E(k)$

\vspace{2ex}
(ii) $\quad \int_0^Kdn^4xdx=\frac{4-2k^2}{3}E(k)+\frac{k^2-1}{3}K(k)$

\vspace{2ex}
(iii) $\quad \int_0^Kdn^6xdx=\frac{8k^4-23k^2+23}{15}E(k)+\frac{4(k^2-1)(2-k^2)}{15}K(k)$

\vspace{2ex}
(iv) $\quad \int_0^Kdn^2xsn^2xdx=\frac{2k^2-1}{3k^2}E(k)+\frac{1-k^2}{3k^2}K(k)$

\vspace{2ex}
(iv) $\quad \int_0^Kdn^2xsn^4xdx=\frac{8k^4-3k^2-2}{15k^4}E(k)+\frac{2(1-k^2)(1+2k^2)}{15k^4}K(k)$

\vspace{2ex}
(v) $\quad \int_0^{2K}cn^2xdx=\frac{2}{k^2}[E(k)+(k^2-1)K(k)]$

\vspace{2ex}
(vi) $\quad \int_0^{2K}sn^2xdx=\frac{2}{k^2}[K(k)-E(k)]$

\vspace{2ex}
(vii) $\quad \int_0^{K}dn xdx= \int_0^{2K}dnx sn^2xdx = \int_0^{2K}dnx cn^2xdx=\pi/2$

\vspace{2ex}
(viii) $\quad \int_0^{2K}dn xsn^2x cn^2xdx=\pi/8$

\vspace{2ex}
(ix) $\quad \int_0^{2\pi}dn^3 xdx=\frac{2-k^2}{2}\pi$

\vspace{2ex}
Now, we first recall that
$$\begin{array}{l}
$$r(x)=\alpha^2[1+k^2+\sqrt{1-k^2+k^4}-3k^2sn^2(\alpha x;k)],\\
\psi_0(x)=dn(x;k)[1-(1+2k^2-\sqrt{1-k^2+4k^4})sn^2(x;k)]>0.
\end{array}$$
From (\ref{2.2}), we obtain after integration
$$||r||^2=w\int_{0}^{T}{r(x)}dx=wr_1T+w(r_0-r_1)\int_{0}^{T}{cn^2(\alpha x)}dx.$$
Then, by using (v) and the expressions of $\omega$, $r_1$ etc, we calculate
$$||r||^2=16\alpha^3\sqrt{1-k^2+k^4}[(k^2-2+\sqrt{1-k^2+k^4})K(k)+3E(k)]$$
Similarly, by direct calculations and making use of (i)-(ix), we come to the
expressions
$$||\psi_0||^2=\frac{4}{\alpha}\frac{\sqrt{1-k^2+4k^4}}{15k^4}
\{(k^2-1)[(8k^4+3k^2+2)-2(2k^2+1)\sqrt{1-k^2+4k^4}]K(k)$$
$$+[(8k^4-3k^2-2)\sqrt{1-k^2+4k^4}-2(8k^6-4k^4-k^2-1)]E(k)\}.$$

$$\langle r,\psi_0\rangle=\pi\alpha[(1+k^2+\sqrt{1-k^2+k^4})
(1-2k^2+\sqrt{1-k^2+4k^4})$$
$$-\frac{9k^2}{4}(\frac13-2k^2+\sqrt{1-k^2+4k^4})]$$

Next, calculating the asymptotical expansions near $k=0$, we obtain
$$\begin{array}{l}
||r||^2=16\pi\alpha^3(1-\frac34k^2+\frac{39}{64}k^4+\ldots)\\[2mm]
\langle r,\psi_0\rangle=4\pi\alpha(1-\frac74k^2+\frac{71}{32}k^4+\ldots)\\[3mm]
||\psi_0||^2={\displaystyle\frac{\pi}{\alpha}}\sqrt{1-k^2+4k^4}(1-\frac94k^2+\frac{135}{64}k^4+\ldots)\\[3mm]
{\displaystyle\frac{-\lambda_0}{\lambda_2-\lambda_0}=
\frac{1-\frac12k^2+\frac{15}{16}k^4+\ldots}{\sqrt{1-k^2+4k^4}}}
\end{array}$$
and, finally,
\begin{equation}\label{4.6}
\frac{-\lambda_0}{\lambda_2-\lambda_0}\frac{||r||^2||\psi_0||^2}
{\langle r,\psi_0\rangle^2}=1-\frac{21}{32}k^4+O(k^6).
\end{equation}
Let us denote by $1-\delta$ the right-hand side of (\ref{4.6}). Clearly,
$\delta=\delta(k)$ satisfies  $1>\delta>0$ for $k$ small enough, say
$0<k\leq k_0$. Fixing such a small $k$, then (\ref{4.6}) yields
$$\frac{||r||^2||\psi_0||^2}
{\langle r,\psi_0\rangle^2}=(1-\delta)
\left(1-\frac{\lambda_2}{\lambda_0}\right)\leq
1+\frac{\lambda_2}{\lambda_0}(\delta-1).$$
Therefore
$$C_1=\lambda_2+\lambda_0\left(1+\frac{1}{d}\right)\left(
\frac{||r||^2||\psi_0||^2}{\langle r,\psi_0\rangle^2}-1\right)\geq
\lambda_2\left[1+\left(1+\frac{1}{d}\right)(\delta-1)\right]=C\lambda_2$$
where $C$ is a positive constant, provided that $d$ is chosen sufficiently
large. Note that the above estimate is not uniform in $k$ when $k$ tends
to zero. This is because $\delta =O(k^4)$ and hence $d=O(k^{-4})$.

When $k\geq k_0$, we can simply draw the graph of the corresponding function
$f(k)=\frac{||r||^2||\psi_0||^2}{\langle r, \psi_0 \rangle^2}-1+\frac{\lambda_2}{\lambda_0}$
to see that it is negative and placed far from zero.
Hence, $C_1\geq C\lambda_2$, too (uniformly for $k\geq k_0$).

 \vspace{1cm}
  \begin{figure}[h]
\centering
\includegraphics[width=5cm,height=5cm]{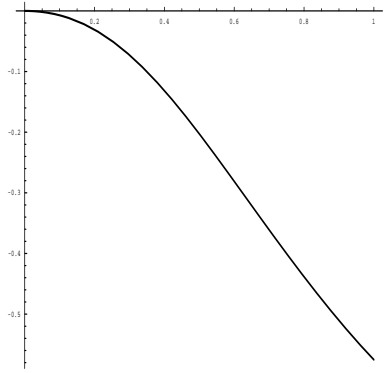}
\caption{Graph of $f(k)<0$}
\label{fig2}
\end{figure}

\vspace{2ex}
\noindent
{\bf D1. The estimates for  $D_{11}=\frac{||r^2-\omega r||}{||r'||}$ and
$D_{12}=\frac{||r'||}{||r||}$.}
Making use of statements (i) and (ii) of the Lemma, we have
$$D_{11}^2=\frac{\int_0^T(r^2-\omega r)^2dt}{\int_0^T(c+\omega r^2-\frac23r^3)dt}=
\frac{\omega^2I_2'-2\omega I_3'+I_4'}{cI_0'+\omega I_2'-\frac23I_3'}$$
$$=\frac57.\frac{-3c\omega  I_0'+(6c+\omega^3)I_1'}{3cI_0'+\omega^2I_1'}
=\frac{-\omega I_0+2I_1}{I_0}\leq 2\omega.$$
The last inequality follows from the fact that the periodic trajectories of the
Hamiltonian system lie inside the saddle loop, in the domain
$H(x,y)<0$ which implies $y^2<x^2(\omega-\frac23x)$ and $x<\frac32\omega$.

Similarly,
$$D_{12}^2=\frac{\int_0^T(c+\omega r^2-\frac23r^3)dt}{\int_0^Tr^2dt}=
\frac{cI_0'+\omega I_2'-\frac23 I_3'}{I_2'}$$
$$=\frac{3c  I_0'+\omega^2 I_1'}{5\omega I_1'}
=\frac{(3c+\omega^3)I_0}{\omega(7I_1-\omega I_0)}\leq
\frac{3c+\omega^3}{5\omega^2}.$$
The last inequality is due to the known fact \cite{Roussarie} (Theorem 12),
that the function $Q(c)=I_1(c)/I_0(c)$
is strictly decreasing, therefore $Q(c)>Q(0)=\frac67\omega$, the value obtained by direct
calculation of elementary integrals.

\vspace{2ex}
  \section{Appendix 2}
%%%%%%%%%%%%%%%%%%%%%%%%%%%%%%%%%%%%%%%%%%%%%%%%%%%%%%%%%%%%%%%%%%%%%%%%%%

First of all, let us mention that the operator $L_1$, defined by differential
expression (\ref{oper}) and equipped with periodic boundary conditions in
$[0,T]$, has the following spectral data
\begin{equation}\label{spec}
\begin{array}{ll}
\lambda_0=-\omega -\sqrt{4\omega^2+6c}, & \psi_0=3r^2-3\omega-\lambda_0,\\
\lambda_1=0, & \psi_1=r',\\
\lambda_2=-\omega +\sqrt{4\omega^2+6c}, & \psi_2=3r^2-3\omega-\lambda_2.
\end{array}
\end{equation}
For $n\in\Z$ and $c\in(-\frac12\omega^2,0)$, consider the
line integrals $I_n(c)$ and their  derivatives $I'_n(c)$ given by
\begin{equation}\label{in}
I_n(c)=\oint_{H=c}x^nydx,\qquad I_n(c)=\oint_{H=c}\frac{x^ndx}{2y}
\end{equation}
where one can assume for definiteness that the integration is along the
right oval contained in the level set $\{H=c\}$, $H(x,y)=y^2-\omega x^2+\frac12x^4$.
As above, these integrals satisfy $I'_n(c)=\frac12\int_0^Tr^n(z)dz$.
The properties of $I_n$ are well known. We only list some facts we are going
to use.

\vspace{2ex}
\noindent
{\bf Lemma.}  (i)  {\it The following identity holds:
$$(n+6)I_{n+3}-2\omega(n+3)I_{n+1}-2ncI_{n-1}=0,\quad n\in\Z$$
which implies}
\begin{equation}\label{recur}
I_3=\omega I_1,\quad
I_4=\frac{2c}{7}I_0+\frac{8\omega}{7}I_2,\quad
I_6=\frac{8\omega c}{21}I_0+\left(\frac{2c}{3}+\frac{32\omega^2}{21}\right)I_2.
\end{equation}

\vspace{1ex}
\noindent
(ii) {\it The integrals $I_0$ and $I_2$ satisfy the system}
$$\begin{array}{l}
4cI'_0+2\omega I'_2=3I_0,\\
4\omega cI'_0+(12c+8\omega^2)I'_2=15I_2.\end{array}$$

\vspace{1ex}
\noindent
(iii) {\it The ratio $R(c)=I_2'(c)/I_0'(c)$ satisfies the Riccati equation
and related system
\begin{equation}\label{ricc}
(8c^2+4\omega^2c)R'(c)=-2\omega c+4cR(c)+\omega R^2(c),\qquad
\begin{array}{l}\dot{c}=8c^2+4\omega^2c,\\
\dot{R}=-2\omega c+4cR+\omega R^2,\end{array}\end{equation}
which imply estimates}
\begin{equation}\label{est}
-\frac{2c}{\omega}\leq R(c)\leq-\frac{c}{2\omega}+\frac{3\omega}{4}.
\end{equation}

\vspace{2ex}
\noindent
The equations in (i)--(iii) are derived in a standard way, see \cite{HIK1}
for more details. The estimates (\ref{est}) follow from the fact that,
in the $(c,R)$-plane,
the graph of $R(c)$ coincides with the concave separatrix trajectory of
the system (\ref{ricc}) contained in the triangle with vertices $(0,0)$,
$(-\frac12\omega^2,\omega)$ and $(0,\frac34\omega)$ and connecting the first
two of them. Note that $R'(c)=-1/(2\omega)$ at $c=-\frac12\omega^2$.
We also use analyticity of the ratio $R(c)$
 at this point and properties of the phase portrait of (\ref{ricc}) to verify
the above statements.

After this preparation, we turn to prove the estimates we used in the
preceding sections.

\vspace{2ex}
\noindent
{\bf A2. The estimate for } $A_2=\frac{||r^2||^2}{||r||^2}$.
By (\ref{red1}), (\ref{recur}) and the first inequality in (\ref{est}),
we have
$$A_2=\frac{\int_0^Tr^4dt}{\int_0^Tr^2dt}=\frac{I_4'}{I_2'}=
\frac{2cI_0'+4\omega I_2'}{3I_2'}=\frac{2c}{3R}+\frac{4\omega}{3}\geq \omega.$$

\vspace{2ex}
\noindent
{\bf B2. The estimate for} $B_2=\frac{||rr'||^2}{||r'||^2}$. As
$\displaystyle I'_6=\frac{16\omega c}{15}I'_0+\left(\frac{6c}{5}+\frac{32\omega^2}{15}\right)I'_2,$
we have by (\ref{3.3}) as above
$$B_2=\frac{\int_0^Tr^2(c+\omega r^2-\frac12r^4)dt}{\int_0^T(c+\omega r^2
-\frac12r^4)dt}
=\frac{2cI_2'+2\omega I_4'-I_6'}{2cI_0'+2\omega I_2'-I_4'}$$
$$=\frac25\frac{\omega cI'_0+(3c+3\omega^2)I'_2}{2cI'_0+\omega I'_2}
=\frac25\frac{\omega c+(3c+2\omega^2)R}{2c+\omega R}\geq\frac{4\omega}{5}.$$
To obtain the last inequality, we used both estimates in (\ref{est}).

\vspace{2ex}
\noindent
{\bf C2. The estimate for}
$C_2=\lambda_2+\lambda_0(1+\frac{1}{d})\frac{||\psi||^2}{\nu^2||\psi_0||^2}$.
By (\ref{3.11}) and (\ref{3.12}) we have
$$C_2=\lambda_2+\lambda_0\left(1+\frac{1}{d}\right)
\left(\frac{||r||^2||\psi_0||^2}{\langle r,\psi_0\rangle^2}-1\right).$$
Next,
$$\langle r,\psi_0\rangle=\int_0^T[3r^3-(3\omega+\lambda_0)r]dt
=6I_3'-(6\omega+2\lambda_0)I'_1=-2\lambda_0I'_1,$$
$$\begin{array}{rl}
||r||^2||\psi_0||^2 &=\int_0^Tr^2dt\int_0^T(3r^2-3\omega-\lambda_0)^2dt\\[4mm]
&=4I_2'[9I'_4-(18\omega+6\lambda_0)I'_2+(3\omega+\lambda_0)^2I_0']\\[2mm]
&=4I_2'[(6c+(3\omega+\lambda_0)^2)I'_0-(6\omega+6\lambda_0)I'_2].
\end{array} $$
By (\ref{3.4}), we have
\begin{equation}\label{EK}
I_2'(c)=\frac12\int_0^Tr^2dt=(r_0^2/\alpha)\int_0^{K(k)}dn^2(t)dt=\sqrt{2}r_0E(k).
\end{equation}
Making use of the identity $E(k)=\frac12\pi F(\frac12,-\frac12,1,k^2)$
where $F$ is the Gauss hypergeometric function, we obtain an appropriate
expansion to estimate $E$ from above
$$
E(k)=\frac{\pi}{2}\left(1-\frac{k^2}{4}-\frac{3k^4}{64}-\frac{5k^6}{512}
-\ldots,\right), \quad
E^2(k)\leq\frac{\pi^2}{4}\left(1-\frac{k^2}{2}-\frac{k^4}{32}\right)$$
with all removed terms negative. As $I_1'=\frac1{\sqrt2}\pi$, by (\ref{3.4})
this implies
$$I_2'^2\leq - I_1'^2\frac{\omega^2-10\omega r_0^2+r_0^4}{8r_0^2}.$$
Together with $I_0'I_2'\geq I_1'^2$, this yields
$$\begin{array}{rl}\displaystyle
\displaystyle \frac{||r||^2||\psi_0||^2}{\langle r,\psi_0\rangle^2}-1
&\displaystyle \leq\frac{1}{\lambda_0^2}\left[6c+(3\omega+\lambda_0)^2
+\frac{3}{4r_0^2}(\omega+\lambda_0)(\omega^2-10\omega r_0^2+r_0^4)\right]-1\\[3mm]
&=\displaystyle\frac{\lambda_2}{\lambda_0}\left(\frac{\lambda_2+\omega}{4r_0^2}-1\right)
\leq\frac{\lambda_2}{\lambda_0}\left(\frac{\sqrt3}{8}-1\right)\end{array}$$
where the equality is obtained by direct calculations.
Therefore,
$$C_2\geq \lambda_2\left(-\frac{1}{d}+\frac{d+1}{d}\frac{\sqrt{3}}{8}\right)
=\bar{C}_2\lambda_2$$
with $\bar{C}_2>0$ an absolute constant when $d\geq 4$ is fixed.

As a by-product of our calculations, we easily obtain also the estimate
$$\frac{\lambda_0}{\nu^2||\psi_0||^2}=
\frac{\lambda_0||\psi_0||^2}{\langle r,\psi_0\rangle^2}
\geq \frac{\lambda_2(\frac{\sqrt3}{8}-1)+\lambda_0}{||r||^2}
\geq -D_1\omega^\frac12.$$

\vspace{2ex}
\noindent
{\bf D2. The estimates for  $D_{21}=\frac{||r^3-\omega r||}{||r'||}$ and
$D_{22}=\frac{||r'||}{||r||}$.}  We proceed as in case {\bf D1} above.
Making use of statements (i) and (ii) of the Lemma, we have
$$D_{21}^2=\frac{\int_0^T(\omega^2r^2-2\omega r^4+r^6)dt}{\int_0^T(c+\omega r^2-\frac12r^4)dt}=
\frac{\omega^2I_2'-2\omega I_4'+I_6'}{cI_0'+\omega I_2'-\frac12 I_4'}$$
$$=\frac{-4c\omega I_0'+(18c+7\omega^2)I_2'}{5(2cI_0'+\omega I_2')}
=\frac{-\omega I_0+3I_2}{I_0}\leq 5\omega.$$
Similarly,
$$D_{22}^2=\frac{\int_0^T(c+\omega r^2-\frac12r^4)dt}{\int_0^Tr^2dt}=
\frac{cI_0'+\omega I_2'-\frac12 I_4'}{I_2'}$$
$$=\frac{2c I'_0+\omega I'_2}{3I'_2}=\frac{(2c+\omega^2)I_0}{5I_2-\omega I_0}\leq \frac{2c+\omega^2}{3\omega}.$$
As before, we used that $I_2(c)/I_0(c)$ is a decreasing function and
calculated the value $I_2(0)/I_0(0)=\frac45\omega$.

\vspace{2ex}
\noindent
{\bf Acknowledgment.} The first author has been partially supported by a 
Research grant DDVU 02/91 (2010) of the Bulgarian Ministry of Education and Science.

 \end{document}